\documentclass[12pt,preprint]{aastex}

\begin{document}

\title{Sensitive VLBI Observations of the $z = 4.7$ QSO BRI~1202--0725}

\author{Emmanuel Momjian}
\affil{NAIC, Arecibo Observatory, HC 3, Box 53995, Arecibo, PR 00612}
\email{emomjian@naic.edu}

\author{Christopher L. Carilli}
\affil{National Radio Astronomy Observatory, P. O. Box O, Socorro, NM, 87801}
\email{ccarilli@nrao.edu}

\author{Andreea O. Petric}
\affil{Astronomy Department, Columbia University, New York, NY, 10027}
\email{andreea@astro.columbia.edu}

\begin{abstract}

We present sensitive phase-referenced VLBI results on the radio continuum emission from the $z=4.7$ double source BRI~1202--0725. The observations were carried out at 1425 MHz using the Very Long Baseline Array (VLBA), the phased Very Large Array (VLA), and the Green Bank Telescope (GBT).

Our sensitive VLBI images of BRI~1202--0725 at $0.25 \times 0.14$~arcsec resolution show a continuum structure in each of its two components. Fitting Gaussian models to these continuum structures yield total flux densities of $315 \pm 38$ and $250 \pm 39~\mu$Jy, for the northern and the southern components, respectively. The estimated intrinsic brightness temperatures of these continuum structures are $\sim 2\times 10^4$~K. Neither component is detected at the full VLBI resolution (29 mas $\times 7$ mas), with a 4$\sigma$ point source upper limit of 40~$\mu$Jy~beam$^{-1}$, or an upper limit to the intrinsic brightness temperature of $6.7\times 10^5$ K. The highest angular resolution with at least a 4$\sigma$ detection is $\sim$85~mas. At this resolution, the images reveal a single continuum feature in the northern component of BRI~1202--0725, and two continuum features in the southern component, separated by 320~mas. This is similar to the structures seen in the high resolution images of the CO emission.

The extent of the observed continuum sources at 1.4~GHz and the derived brightness temperatures are consistent with nuclear starbursts. Moreover, the absence of any compact high-brightness temperature source suggests that there is no radio-loud AGN in BRI~1202--0725.
\end{abstract}


\section{INTRODUCTION}

Optical surveys such as the Sloan Digital Sky Survey (SDSS; York et al.~2000) and the Digitized Palomar Sky Survey \citep{DJO99} have revealed large samples of quasi-stellar objects out to $z \sim 6$. Studies by \citet{FAN02} have shown that at such a high redshift we are approaching the epoch of reionization, the edge of the ``dark ages'', when the first stars and massive black holes were formed.

Observations of high redshift QSOs at mm and sub-mm wavelengths have shown that a significant fraction ($\sim 30\%$) of the sources are strong emitters of far-IR radiation, with luminosities $L_{\rm FIR} > 10^{12}~L_{\odot}$, and molecular gas masses greater than $10^{10}~M_{\odot}$ \citep{BV91,BAR94,OHT96,OMO96b,GUI97,FRA99,OMO01,CAR02,CAR04}. Moreover, the large reservoirs of warm gas and dust in these objects have led to the hypothesis that these are starburst galaxies with massive star formation rates on the order of 1000~$M_{\odot}~{\rm yr}^{-1}$ \citep{HD99,LIL99}.

An important question regarding these high-$z$ QSOs is whether the dominant dust heating mechanism is an AGN or a starburst. The high resolution of Very Long Baseline Interferometry (VLBI) observations permits a detailed look at the physical structures in the most distant cosmic sources. Also, sensitive VLBI continuum observations can be used to determine the nature of the energy source(s) in these galaxies at radio frequencies.

To date, several high redshift QSOs have been imaged at milliarcsecond resolution \citep{FRE97,FRE03,MOM03,BEE04}. In this paper, we present sensitive VLBI observations of the $z=4.7$ QSO BRI~1202--0725. This object was identified in the Automatic Plate Measuring survey \citep{IMH91}, and its optical spectra show strong associated Ly$\alpha$ absorption \citep{STO96}. BRI~1202--0725 was the first high redshift QSO in which thermal dust emission was detected \citep{MC94},

Continuum emission at centimeter and millimeter wavelengths, as well as CO spectral line observations, showed that BRI~1202--0725 is a double source with an angular separation of $4''$. The total far-IR luminosity of this galaxy is $L_{\rm FIR}=4.2 \times 10^{13}~L_{\odot}$, with a dust temperature of $\sim70$~K \citep{OHT96,OMO96a,OMO96b,LMA01,CAR02}.

Various studies suggest that the observed double source may indicate a pair of interacting objects separated by 28~kpc, and that the large apparent luminosity of BRI~1202--0725 may not be due to gravitational lensing \citep{OMO96b,YUN00,CAR02}. Supporting this argument is the clear difference in the CO line widths and redshifts of the northern and southern components \citep{OMO96b,CAR02}.  Moreover, optical imaging of BRI~1202--0725 show a single point-like QSO with $M_{\rm B}=-28.5$, with faint near-IR and Ly$\alpha$ emission extending about $2\rlap{.}^{''}4$ north of the QSO \citep{HME96}. The optical QSO is associated with the southern component seen in CO line, centimeter, and millimeter, observations \citep{CAR02}. Most recently, \citet{KLAM04} have proposed that the 4$''$ double structure in BRI 1202--0725 could be due to jet induced star formation, with the northern component corresponding to a radio `hot spot' (i.~e. jet-shock or terminus).

The CO~($2-1$) emission results with the VLA at $0\rlap{.}''3 \times 0\rlap{.}''21$ resolution \citep {CAR02} reveal a marginally detected source in the northern component of BRI~1202--0725 with an extent of $\geq 0\rlap{.}''5$. The southern component of BRI~1202--0725 appears as two unresolved sources separated by $0\rlap{.}''3$ in these CO~($2-1$) observations.

Throughout this paper, we assume a flat cosmological model with $\Omega_{m}=0.3$, $\Omega_\Lambda=0.7$, and ${H_{0}=65}$~km~s$^{-1}$~Mpc$^{-1}$. In this model, at the distance of BRI 1202--0725, 1~milliarcsecond (mas) corresponds to 7~pc. We also define the spectral index $\alpha$, as a function of frequency ($\nu$), to be $S_{\nu}\propto \nu^{\alpha}$.

\section{OBSERVATIONS AND DATA REDUCTION}

The VLBI observations were carried out at 1.4~GHz on 2003 April 16 and 19, using the Very Long Baseline Array (VLBA), the phased Very Large Array (VLA), and the Green Bank Telescope (GBT) of the NRAO\footnote{The National Radio Astronomy Observatory is a facility of the National Science Foundation operated under cooperative agreement by Associated Universities, Inc.}. Four adjacent 8~MHz baseband channel pairs were used in the observations, both with right and left-hand circular polarizations, and sampled at two bits. The data were correlated at the VLBA correlator in Socorro, NM, with 2~s correlator integration time and two independent correlation passes for the two components of BRI~1202--0725, which are separated by $4''$. The total observing time was 14~hr. Table~1 summarizes the parameters of these observations.
 
The observations employed nodding-style phase referencing, using the calibrator J1202--0528, with a cycle time of 4~min, 3~min on the target source and 1~min on the calibrator. A number of test cycles were also included to monitor the coherence of the phase referencing. These tests involved switching between two calibrators, the phase calibrator J1202--0528 and the phase-check calibrator J1152--0841, using a similar cycle time to that used for the target source.

The accuracy of the phase calibrator position is important in phase-referencing observations \citep {WAL99}, as this determines the accuracy of the absolute position of the target source and any associated components. Phase referencing, as used here, is known to preserve absolute astrometric positions to better than $\pm 0.01''$ \citep{FOM99}.

Data reduction and analysis were performed using the Astronomical Image Processing System (AIPS) and Astronomical Information Processing System (AIPS$++$) of the NRAO.  After applying {\it a priori} flagging, amplitude calibration was performed using measurements of the antenna gain and system temperature for each station. Ionospheric corrections were applied using the AIPS task ``TECOR''. The phase calibrator J1202--0528 was self-calibrated in both phase and amplitude and imaged in an iterative cycle.

Images of the phase-check calibrator, J1152--0841, were deconvolved using two different approaches: (a) by applying the phase and the amplitude self-calibration solutions of the phase reference source J1202--0528 (Figure~1{\it{a}}), and (b) by self calibrating J1152--0841 itself, in both phase and amplitude (Figure~1{\it{b}}). The peak surface brightness ratio of the final images from the two approaches gives a measure of the effect of residual phase errors after phase referencing (i.e. `the coherence' due to phase referencing). At all times, the coherence was found to be better than 97\%.

The self-calibration solutions of the phase calibrator, J1202--0528, were applied to the two correlation passes of the target source, BRI~1202--0725. Each of the two components of the target source was then deconvolved and imaged at various spatial resolutions by tapering the visibility data.

In addition to these VLBI observations, 1.4~GHz VLA A-array (resolution $\sim 1.5''$) observations carried out in August 1999 and November 2004 were reduced and analyzed for comparison purposes. 

\section{RESULTS \& ANALYSIS}

Figures 2 and 3 are naturally weighted images of the northern and southern components of BRI~1202--0725 at 1.4~GHz with a resolution of $252 \times 141$~mas (P.~A.= $-53^{\circ}$). The images were obtained by applying a two-dimensional Gaussian taper falling to 30\% at 1~M$\lambda$ in both the {\it u}- and {\it v}-directions of the visibility data. The rms noises in these images are 28.3 and $26.8~\mu$Jy~beam$^{-1}$ for the northern and southern components, respectively.

Table~2 shows the results of fitting Gaussian models to the observed source spatial profiles using the AIPS++ task ``IMAGEFITTER''. The source names, redshifts, and positions are listed in columns 1, 2, 3, and 4, respectively. Columns 5 and 6 show the peak and total flux densities of the fitted Gaussian functions. Columns 7, 8, and 9 list the fitted half-power ellipse axes and the position angles.  The intrinsic brightness temperatures (corresponding to a rest-frame frequency of 8 GHz) of the radio continuum structures seen in the northern and southern components of BRI~1202--0725 are $(2.2 \pm 0.3) \times 10^4$ and $(1.7 \pm 0.3) \times 10^4$~K, respectively. The total radio flux density of BRI~1202--0725 at 1.4~GHz, i.~e., the sum of the flux densities of both components, is $565 \pm 55~\mu$Jy, as measured with our sensitive VLBI array.

No continuum emission is detected at the the full resolution of the array, which is $29 \times 7$~mas (PA=$-6^{\circ}$), indicating the absence of any radio continuum emission with flux densities of $\geq 4\sigma \simeq 40~\mu$Jy~beam$^{-1}$. This implies an upper limit to the intrinsic brightness temperature of $6.7 \times 10^5$~K for any compact radio source in BRI~1202--0725. We emphasize again that our coherence tests during these observations using two VLBI calibrators show that the lack of a strong point source in BRI~1202--0725 at the full resolution of the array cannot be due to the phase referencing procedure.

Figures 4 and 5 show the two components in BRI~1202--0725 at the highest angular resolutions for which there are at least 4$\sigma$ detections. These detections correspond to resolutions of $81 \times 57$~mas in position angle $-11^{\circ}$ for the northern component (Figure~4; $\sigma=25.7~\mu$Jy~beam$^{-1}$), and $111 \times 83$~mas in position angle $-37^{\circ}$ for the southern component (Figure~5; $\sigma=25.5~\mu$Jy~beam$^{-1}$). These images were obtained by applying two-dimensional Gaussian tapers falling to 30\% at 2.6 and 1.8~M$\lambda$ in both the {\it u}- and {\it v}-directions of the visibility data, for the northern and southern components, respectively.

The intermediate angular resolution image of the northern component (Figure~4) shows a single continuum structure at $4.3\sigma$ level. However, the southern component (Figure~5) seems to be resolved into two continuum structures separated by 320~mas. The stronger continuum source in Figure~5 is detected at $4.8\sigma$ level, and the weaker at $3.7\sigma$. The derived intrinsic brightness temperature of the continuum structure in Figure~4 (northern component) is $(7.5 \pm 1.9) \times 10^4$, and of the structures in Figure~5 (southern component) are $(4.5 \pm 1.0) \times 10^4$ and $(3.5 \pm 1.0) \times 10^4$~K. The extent of these continuum sources range between 0.4 and 1~kpc on the plane of the sky. The single, resolved structure of the northern component, and the double structure of the southern component seen in these VLBI images are similar to the structures seen in the CO~($2-1)$ emission observations of BRI~1202--0725 \citep{CAR02}. In particular, the CO~($2-1)$ emission from the southern component shows a 300 mas separation double morphology along the same position angle as that seen in the radio continuum image of Figure~5.

In addition to the VLBI results, we analyzed VLA A-array data from two epochs, August 1999 and November 2004. The VLA flux densities, which are listed in Table~3, are comparable to the VLBI measurements reported in Table~2.

\section{DISCUSSION}

At a moderate angular resolution ($252 \times 141$~mas; Figures 2 \& 3), the two components of the high $z$ QSO BRI~1202--0725 show radio continuum emission at the observed 1.4 GHz, which corresponds to a rest frequency of 8~GHz at $z=4.7$. The northern and southern components constitute 56 and 44\% of the total flux density of BRI~1202--0725, respectively.
The intrinsic brightness temperature values of these continuum structures are on the order of $2\times 10^4$~K.

At the full resolution of our array ($29 \times 7$~mas), the two components of BRI~1202--0725 are resolved out and do not show any single dominant source of very high brightness temperature ($< 6.7\times 10^5$ K). This is in contrast to the results obtained by \citet{MOM04} on a sample of three high-$z$ radio-loud quasars which were imaged with the VLBA. In each of these $z>4$ quasars, a radio-loud AGN dominates the emission at 1.4~GHz on a few mas size scale, with intrinsic brightness temperatures in excess of $10^9$~K. 

\citet{CON91} have derived an empirical upper limit to the brightness temperature for nuclear starbursts of order $10^5$~K at 8~GHz, while typical radio-loud AGN have brightness temperatures exceeding this value by at least two orders of magnitude. These authors also present a possible physical model for this limit involving a mixed non-thermal and thermal radio emitting (and absorbing) plasma, constrained by the radio-to-FIR correlation for star-forming galaxies. 

For the two components of BRI~1202--0725, the derived intrinsic brightness temperatures from our VLBI observations are typical of starburst galaxies. Also, the linear extents of the radio continuum emission regions in BRI~1202--0725 are typical of local starburst powered Ultra-Luminous IR galaxies (ULIRGs; Sanders \& Mirabel 1996), such as Arp~220, Mrk~273, and IRAS~17208--0014 \citep{SMI98,CAR00,MOM03}.
Moreover, the radio continuum morphology of each component in BRI~1202--0725 is similar to that seen in high resolution images of the CO emission \citep{CAR02}. Co-spatiality of the radio continuum emission with the molecular gas (and presumably dust) is expected for star forming galaxies \citep{CAR03}.

The physical characteristics of the QSO BRI~1202--0725 are also similar to one of the most luminous IR sources, J1409+5628, located at $z=2.58$ \citep{BEE04}. High angular resolution observations at 1.4~GHz showed that the radio continuum emission in this optically very bright QSO is dominated by an extreme nuclear starburst, with a massive star formation rate of $(3-8) \times 10^{3}~M_{\odot}~{\rm yr}^{-1}$.

Another argument in favor of a massive starburst origin for the radio continuum emission from BRI~1202--0725 is the spectral index between the 1.4 and 350~GHz ($\alpha\rlap{$_{1.4}$}^{350}$) for the overall flux density of its two components.
This spectral index, $\alpha\rlap{$_{1.4}$}^{350}$, corresponds to that measured between 
the non-thermal synchrotron radiation at 1.4~GHz and the thermal emission from warm dust at 350~GHz. For BRI~1202--0725, $\alpha\rlap{$_{1.4}$}^{350}$ = $+0.96 \pm 0.04$, and the predicted value for a starburst galaxy at $z=4.7$ is $\alpha\rlap{$_{1.4}$}^{350}=+1.03 \pm 0.10$ \citep{YUN00,CY00}.

\citet{YUN00} suggested possible flux density variation at 1.4 GHz for BRI~1202--0725. Such variation would be hard to reconcile physically with the kpc-scale spatial extent observed with VLBI for the radio components. However, the variability conclusion was based on comparison of VLA observations made with radically different resolutions (A through D array; or $1{''}$ and $60{''}$). We have reexamined the variability using observations made with the VLA A-array in August 1999 and November 2004. The total flux densities of each component in BRI~1202--0725, as reported in Section 3, do not show any sign of variability during a five year period.
Moreover, the measured flux densities with the VLA A-array are consistent with our VLBI results (Table~2), suggesting that the radio continuum emission at 1.4~GHz is confined within the extent of the structures seen in the VLBI images.

While the measured 1.4~GHz radio flux densities in BRI~1202--0725 are consistent with extreme nuclear starburst activity, detecting individual supernovae (SNe) at $z =4.7$ is unlikely.
\citet{SMI98} reported the detection of several luminous radio SNe in the prototype ULIRG Arp~220 with flux densities between 0.2 and 1.17 mJy and angular sizes on the order of 0.25~mas. At the distance of BRI~1202--0725, the flux densities of such luminous radio SNe would be between $ (0.7-4) \times 10^{-3}~\mu$Jy. These values are four orders of magnitude lower that the rms noise levels achieved in our VLBI observations.

In general, the single source morphology of the northern component and the double source morphology of the southern component in BRI~1202--0725, and their starburst nature, lead us to conclude that the northern component could be similar in nature to single nucleus ULIRGs, such as IRAS~17208--0014 \citep{MOM03}, and the southern component to double nucleus ULIRGs, such as Arp~220 \citep{SMI98}. Moreover, these two extreme starburst galaxies (i.~e.~the two components of BRI~1202--0725), that are separated by $4{''}$, may actually be interacting as suggested by various studies \citep{OMO96b,YUN00,CAR02}.

\section{ACKNOWLEDGMENTS}

The authors thanks A. Deshpande, J. Lazio, and C. Salter, for helpful discussions and comments. This research has made use of the NASA/IPAC Extragalactic Database (NED) which is operated by the Jet Propulsion Laboratory, California Institute of Technology, under contract with the National Aeronautics and Space Administration. The Arecibo Observatory is part of the National Astronomy and Ionosphere Center, which is operated by Cornell University under a cooperative agreement with the National Science Foundation.

\clearpage
\begin{figure}
\epsscale{0.70}
\plotone{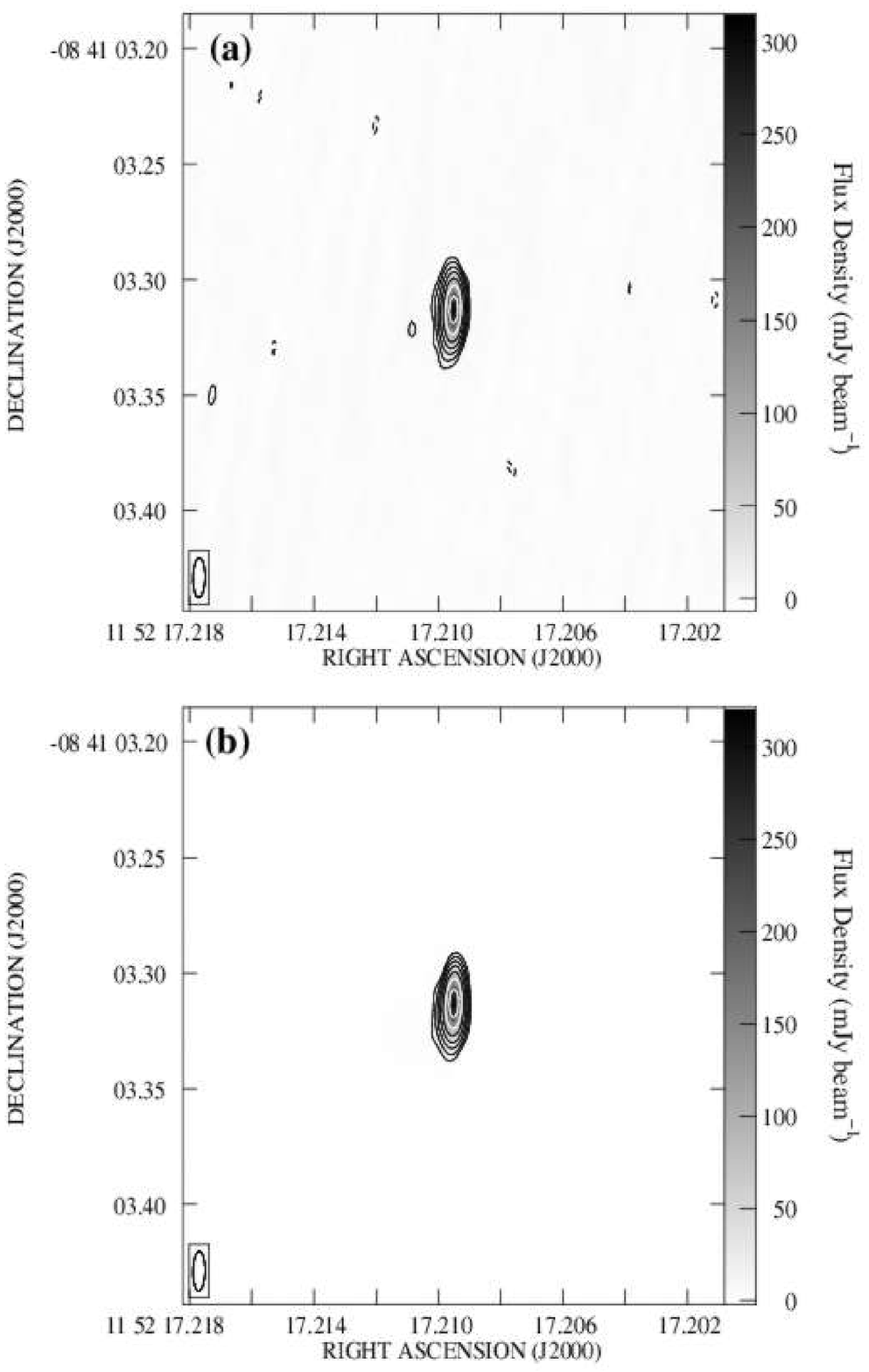}
\figcaption[Momjian.fig1.ps] {Continuum images of the phase-check
calibrator J1152--0841 at 1.4~GHz: a) obtained by applying the phase
and the amplitude self-calibration solutions of the phase reference
source J1202--0528, b) obtained by self calibrating J1152--0841
itself, in both phase and amplitude. The restoring beam size in both
images is $17.2 \times 5.8$~mas in position angle $-1^{\circ}$. The
contour levels are at $-3$, 3, 6, 12, $\ldots$, 192 times the rms
noise level in the phase-referenced image (top), which is
1.2~mJy~beam$^{-1}$. The gray-scale range is indicated by the step
wedge at the right side of each image.
\label{f1}}
\end{figure}

\clearpage
\begin{figure}
\epsscale{0.8}
\plotone{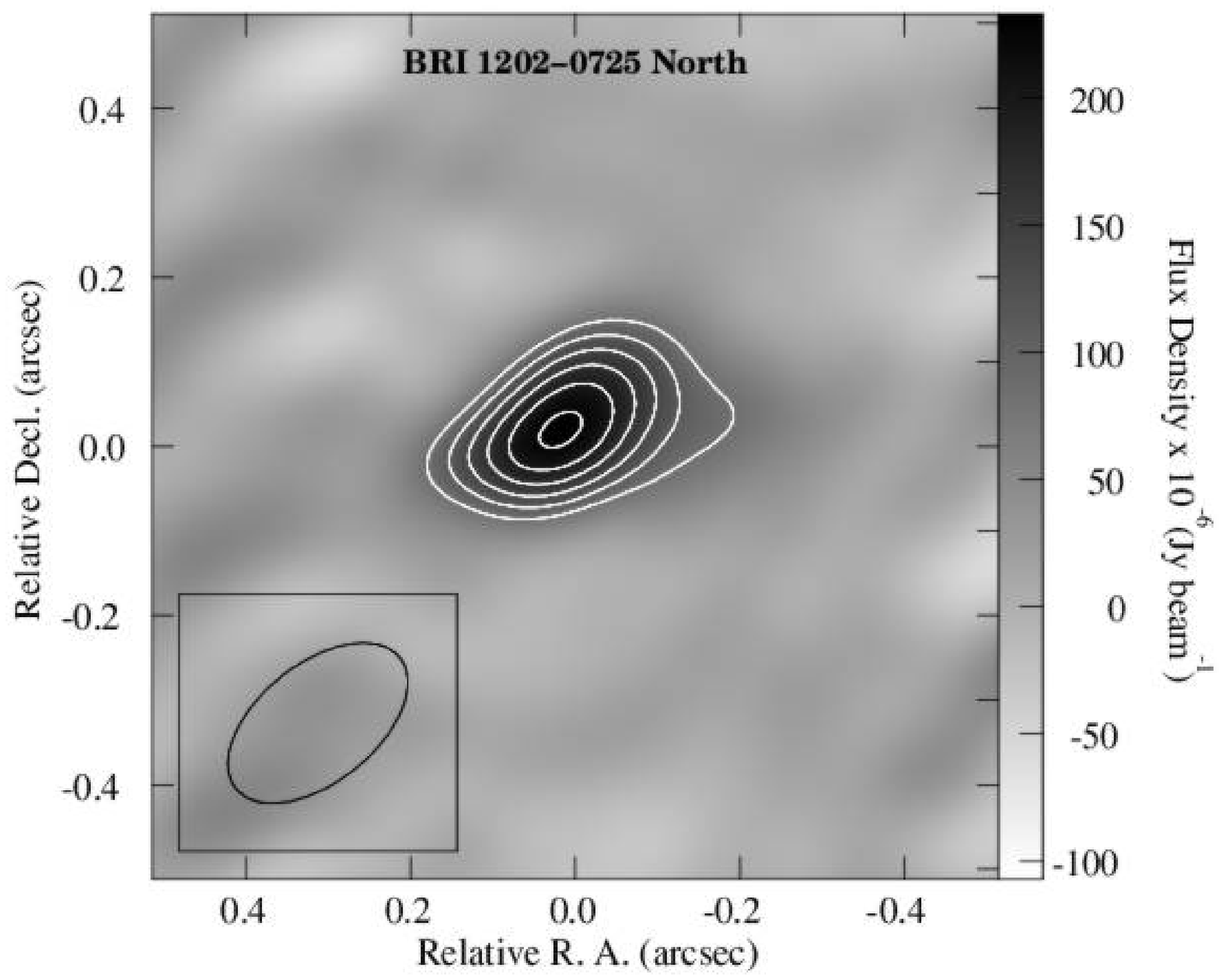}
\figcaption[Momjian.fig2.ps] {Naturally weighted 1.4~GHz continuum
image of the northern component of BRI~1202--0725 at $252 \times
141$~mas resolution (P.~A.=$-53^{\circ}$).  The peak flux density is
232~$\mu$Jy~beam$^{-1}$, and the contour levels are at $-3$, 3, 4, 5,
6, 7, 8 times the rms noise level, which is
28.3~$\mu$Jy~beam$^{-1}$. The gray-scale range is indicated by the
step wedge at the right side of the image. The reference point (0, 0)
is $\alpha(\rm{J2000.0})= 12^{\rm h}05^{\rm m}23\rlap{.}^{\rm s}9785$,
$\delta(\rm{J2000.0})=-07^{\circ}42^{'}29\rlap{.}^{''}7758$.  A two
dimensional Gaussian taper falling to 30\% at 1~M$\lambda$ was
applied.  \label{f2}}
\end{figure}

\clearpage
\begin{figure}
\epsscale{0.8}
\plotone{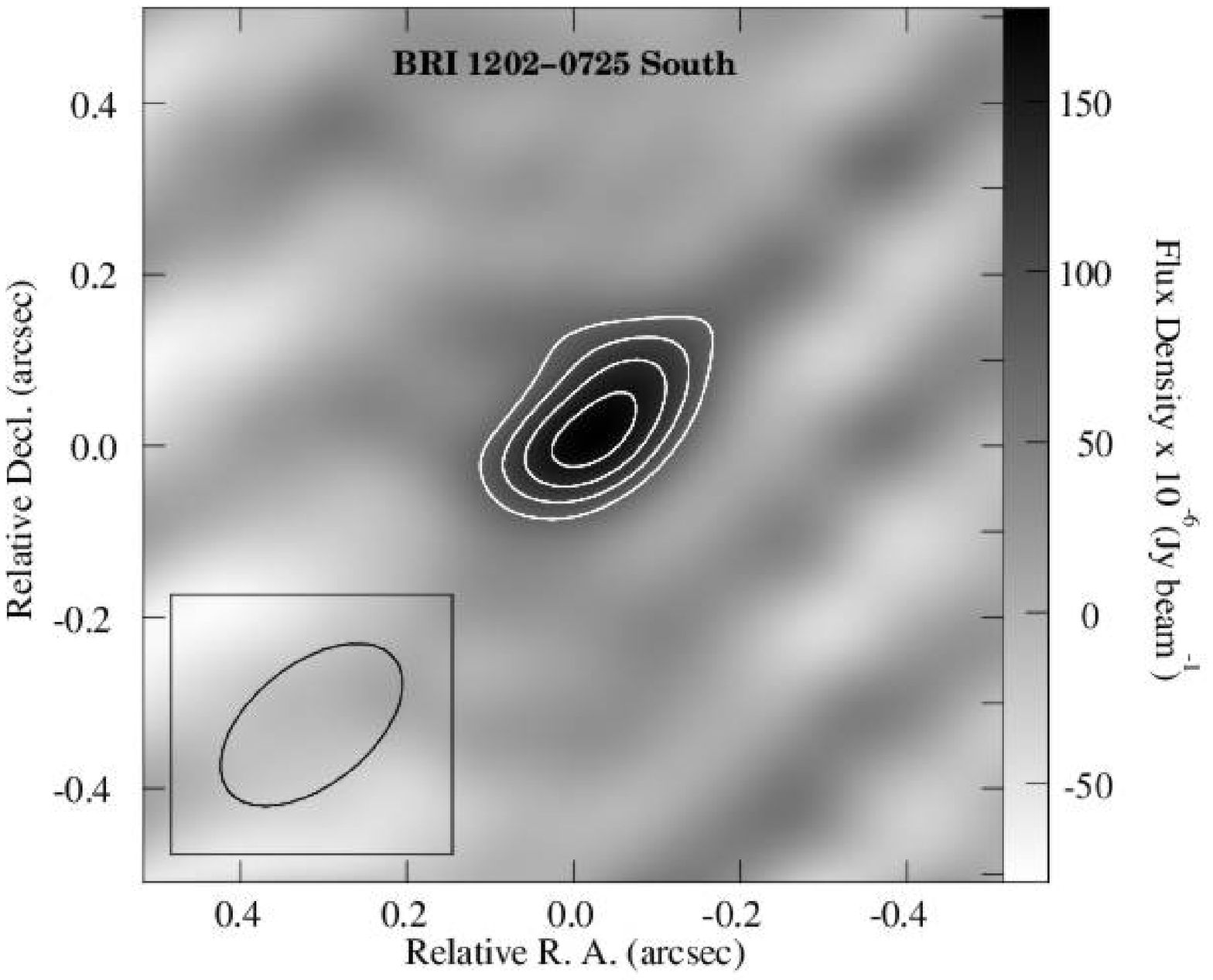}
\figcaption[Momjian.fig3.ps] {Naturally weighted 1.4~GHz continuum
image of the southern component of BRI~1202--0725 at $252 \times
141$~mas resolution (P.~A.=$-53^{\circ}$). The peak flux density is
177~$\mu$Jy~beam$^{-1}$, and the contour levels are at $-3$, 3, 4, 5,
6 times the rms noise level, which is 26.8~$\mu$Jy~beam$^{-1}$. The
gray-scale range is indicated by the step wedge at the right side of
the image.  The reference point (0, 0) is
$\alpha(\rm{J2000.0})=12^{\rm h}05^{\rm m}23\rlap{.}^{\rm s}1187$,
$\delta(\rm{J2000.0})=-07^{\circ}42^{'}33\rlap{.}^{''}1391$.  A two
dimensional Gaussian taper falling to 30\% at 1~M$\lambda$ was
applied.  \label{f3}}
\end{figure}

\clearpage
\begin{figure}
\epsscale{0.8}
\plotone{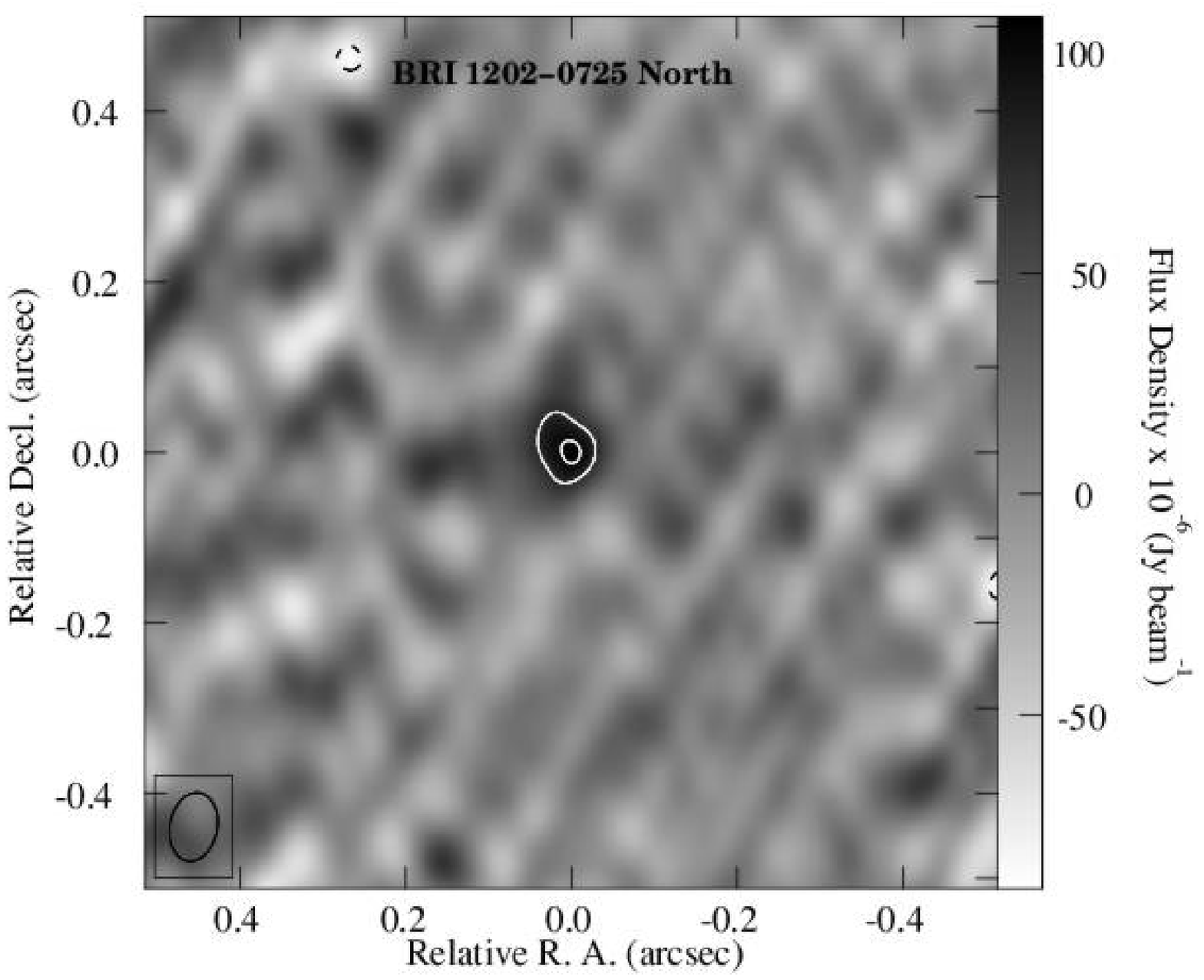}
\figcaption[Momjian.fig6.ps] {Naturally weighted 1.4~GHz continuum
image of the northern component of BRI~1202--0725 at $81 \times
57$~mas resolution (P.~A.=$-11^{\circ}$).  The peak flux density is
108~$\mu$Jy~beam$^{-1}$, and the contour levels are at $-3$, 3, 4
times the rms noise level, which is 25.7~$\mu$Jy~beam$^{-1}$. The
gray-scale range is indicated by the step wedge at the right side of
the image.  The reference point (0, 0) is $\alpha(\rm{J2000.0})=
12^{\rm h}05^{\rm m}23\rlap{.}^{\rm s}9785$,
$\delta(\rm{J2000.0})=-07^{\circ}42^{'}29\rlap{.}^{''}7758$. A two
dimensional Gaussian taper falling to 30\% at 2.6~M$\lambda$ was
applied.  \label{f4}}
\end{figure}

\clearpage
\begin{figure}
\epsscale{0.8}
\plotone{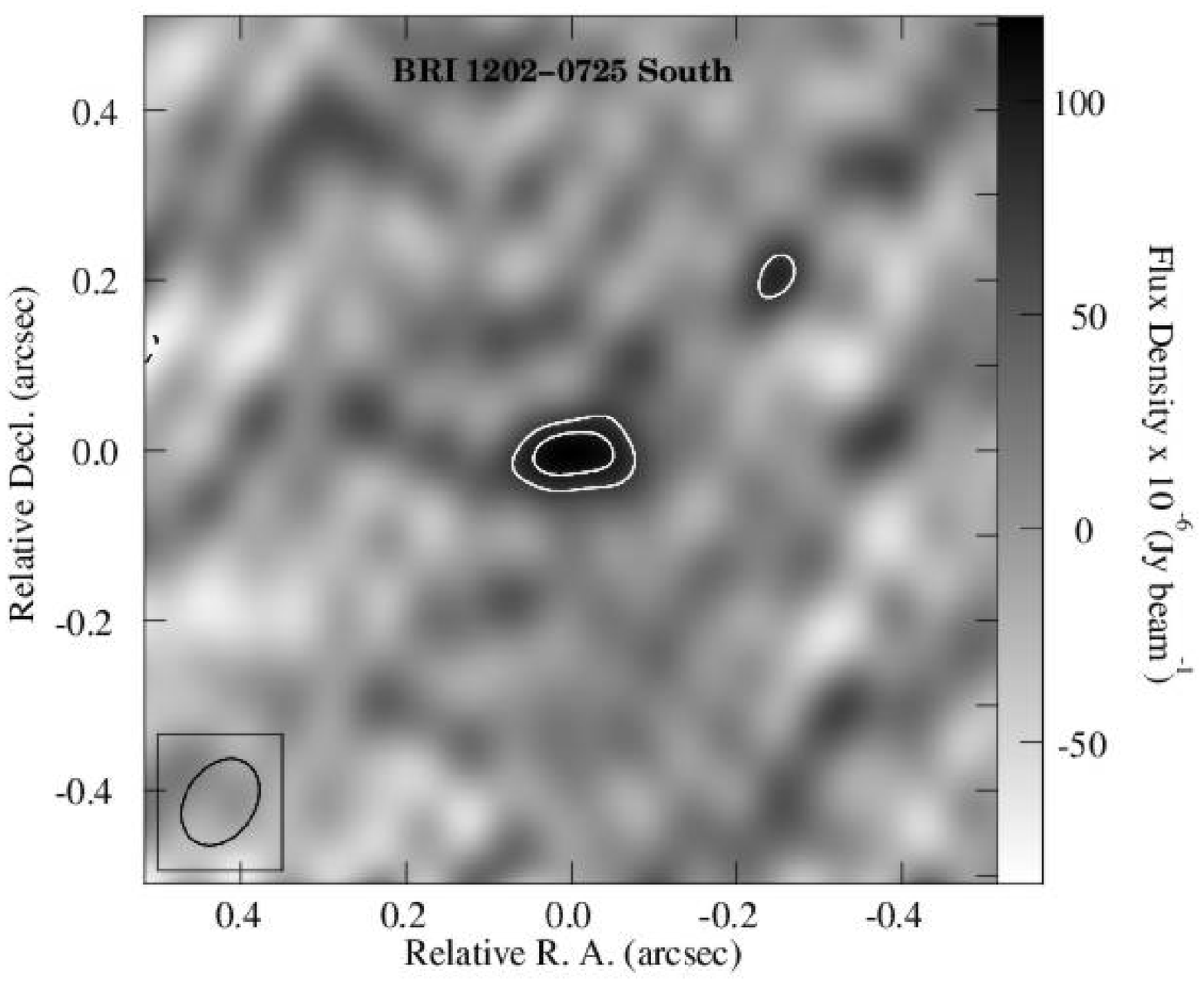}
\figcaption[Momjian.fig7.ps] {Naturally weighted 1.4~GHz continuum
image of the southern component of BRI~1202--0725 at $111 \times
83$~mas resolution (P.~A.=$-37^{\circ}$).  The peak flux density is
110~$\mu$Jy~beam$^{-1}$, and the contour levels are at $-3$, 3, 4
times the rms noise level, which is 25.5~$\mu$Jy~beam$^{-1}$. The
gray-scale range is indicated by the step wedge at the right side of
the image.  The reference point (0, 0) is $\alpha(\rm{J2000.0})=
12^{\rm h}05^{\rm m}23\rlap{.}^{\rm s}1187$,
$\delta(\rm{J2000.0})=-07^{\circ}42^{'}33\rlap{.}^{''}1391$.  A two
dimensional Gaussian taper falling to 30\% at 1.8~M$\lambda$ was
applied.  \label{f5}}
\end{figure}

\clearpage
\oddsidemargin=-1cm
\tabletypesize{\scriptsize}

\begin{deluxetable}{lc}
\tablenum{1}
\tablecolumns{6}\tablewidth{0pc}
\tablecaption{P{\footnotesize ARAMETERS} {\footnotesize OF THE} VLBI O{\footnotesize BSERVATIONS} {\footnotesize OF} BRI~1202--0725}
\tablehead{\colhead{Parameters} & \colhead{Values}}
\startdata
Observing Date \dotfill  & 2003 Apr 16 \& 19 \\
Total observing time (hr)\dotfill  & 14 \\
Phase calibrator\dotfill  & J1202--0528 \\
Phase-referencing cycle time (min)\dotfill  &  4 \\
Frequency (MHz)\dotfill  &  1425 \\
Total bandwidth (MHz)\dotfill   & 32\\
\enddata
\end{deluxetable}
\clearpage
\begin{deluxetable}{ccccccccc}
\tablenum{2}
\tablecolumns{9}
\tablewidth{0pc}
\tablecaption{S{\footnotesize OURCE} P{\footnotesize ARAMETERS} in F{\footnotesize IGURES} 2 \& 3}
\tablehead{
\colhead{}    &\colhead{}
&\multicolumn{6}{c}{Gaussian Component Parameters} \\ \cline{3-9} \\
\colhead{Source} & \colhead{$z$} &\colhead{R.A. (J2000)}   &\colhead{Decl. (J2000)}&
\colhead{Peak} &\colhead{Total} & \colhead{Major Axis\tablenotemark{a,}\tablenotemark{b}} & \colhead{Minor Axis\tablenotemark{a,}\tablenotemark{b}}& \colhead{P. A.\tablenotemark{c}}\\
\colhead{} & \colhead{}  &\colhead{(h m s)}   & \colhead{(${^\circ}~'~''$)}
& \colhead{($\mu$Jy~beam$^{-1}$)} & \colhead{($\mu$Jy)}& \colhead{(mas)}& \colhead{(mas)}& \colhead{(${^\circ}$)} \\
\colhead{(1)}  & \colhead{(2)} & \colhead{(3)}   & \colhead{(4)} &
\colhead{(5)} & \colhead{(6)} &\colhead{(7)} &\colhead{(8)} &\colhead{(9)}}
\startdata
Northern  & 4.6916  & 12 05 22.977   &--07 42 29.750 & $234 \pm  28$& $315 \pm  38$ & 288.6 & 166.5 & 115 \\
Southern  & 4.6947  & 12 05 23.1167   &--07 42 33.109 & $176 \pm  27$& $250 \pm  39$ & 297.4 & 170.6& $130 $\\
\enddata
\tablenotetext{a}{At half maximum.}
\tablenotetext{b}{Gaussian fit error is $\leq$ 0.3~mas.}
\tablenotetext{c}{Gaussian fit error is $\leq$ 0.001${^\circ}$.}
\end{deluxetable}
\clearpage
\begin{deluxetable}{ccccccccc}
\tablenum{3}
\tablecolumns{3}
\tablewidth{0pc}
\tablecaption{M{\footnotesize EASURED} 1.4~GHz F{\footnotesize LUX} D{\footnotesize ENSITIES}\ {\footnotesize WITH THE} VLA A-A{\footnotesize RRAY}}
\tablehead{
\colhead{Source} & \colhead{Date} &\colhead{$S_{\rm 1.4~GHz}$}\\
\colhead{} & \colhead{}  &\colhead{($\mu$Jy)}}
\startdata
Northern  & Aug. 1999   &  $365 \pm 27$ \\
          & Nov. 2004   &  $282 \pm 48$ \\
\tableline 
Southern  & Aug. 1999   &  $230 \pm 34$ \\
          & Nov. 2004   &  $203 \pm 26$ \\
\enddata
\end{deluxetable}

\end{document}